\documentclass{aastex}
\usepackage{spr-astr-addons}
\usepackage{url}
\usepackage{epstopdf}

\RequirePackage{color}

\begin{document}

\title{Colour asymmetry between galaxies with clockwise and counterclockwise handedness}
\shorttitle{Colour asymmetry}
\shortauthors{Lior Shamir}

\author{Lior Shamir\altaffilmark{1}}
\affil{Lawrence Technological University, 21000 W Ten Mile Rd., Southfield, MI 48075, USA. Email: lshamir@mtu.edu}


\begin{abstract}
Recent studies have shown that SDSS galaxies with clockwise patterns are photometrically different from galaxies with anti-clockwise patterns. The purpose of this study is to identify possible differences between the colour of galaxies with clockwise handedness and the colour of galaxies with anti-clockwise handedness. A dataset of 162,514 SDSS galaxies was separated into clockwise and counterclockwise galaxies, and the colours of spiral galaxies with clockwise handedness were compared to the colour of spiral galaxies with anti-clockwise handedness. The results show that the i-r colour in clockwise galaxies in SDSS is significantly higher compared to anti-clockwise SDSS galaxies. The colour difference is strongest between the right ascension of 30$^o$ and 60$^o$, while the RA range of 180$^o$ to 210$^o$ shows a much smaller difference. 
\end{abstract}


\section{Introduction}
\label{introduction}

The handedness of a spiral galaxy is a clear visual characteristic that separates spiral galaxies into galaxies with clockwise handedness and galaxies with counterclockwise handedness. The handedness of a galaxy is subjective to the position of the observer, so that a galaxy with clockwise handedness might seem to have counterclockwise handedness to an observer positioned elsewhere in the universe. Therefore, in a large population of galaxies no differences are expected between the physical properties of galaxies with counterclockwise handedness and the physical properties of galaxies with clockwise handedness.

While some differences between the photometry of clockwise and anti-clockwise galaxies were observed with marginal statistical significance \citep{shamir2013color}, recent experiments using machine learning and statistical methods show strong evidence of photometric differences between clockwise and anti-clockwise spiral galaxies \citep{sha16}. 

These photometric differences were shown with two datasets \citep{sha16}. The first consisted of 13,440 classified manually as spiral by Galaxy Zoo 2 \citep{willett2013galaxy} crowdsourcing campaign, and the second was based on 10,281 galaxies analyzed in a fully automatic process \citep{kum16}, and without any human intervention that could induce human perception bias. Using several different machine learning algorithms to predict the handedness of a galaxy by its photometry, the algorithms correctly predicted the handedness in $\sim$64\% and $\sim$65\% accuracy using the Galaxy Zoo 2 and the automatically classified galaxies, respectively \citep{sha16}.

This paper uses a large set of 162,514 spiral galaxies to show statistically significant differences between the colour of clockwise galaxies and anti-clockwise galaxies imaged by the Sloan Digital Sky Survey (SDSS).

 \section{Data}
\label{computer_dataset}

To obtain a large dataset of galaxies separated to clockwise and anti-clockwise handedness, a catalogue of automatically annotated SDSS DR8 galaxies \citep{kum16} was used. The catalogue contains the annotations of the broad morphology of $\sim$3,000,000 galaxies annotated by an image analysis algorithm \citep{kuminski2014combining}, such that each annotation to spiral or elliptical galaxy is provided with a degree of certainty that the annotation is correct \citep{kum16}. All galaxies annotated as spiral galaxy with certainty of 0.54 or higher were selected to provide a dataset of 740,908 spiral galaxies. Comparison to Galaxy Zoo \citep{lintott2011galaxy} manual classifications shows that galaxies classified as spiral in certainty of 0.54 or higher are in $\sim$98\% of the cases aligned with the ``superclean'' Galaxy Zoo annotations \citep{kum16}.  

As was done in \citep{sha16}, the spiral galaxies were classified into clockwise and counterclockwise galaxies by applying the {\it Ganalyzer} algorithm \citep{shamir2011ganalyzer,ganalyzer_ascl}. Ganalyzer first transforms each galaxy image into its radial intensity plot, and then detects the peaks in each horizontal line of the plot. The peaks are grouped such that each set of peaks can be associated with a galaxy arm. For each arm, the X position of each peak is compared to the X position of the peak in the next line of the plot. If the peaks shift to the left the galaxy is considered a clockwise galaxy, and if the peaks shift to the right the galaxy is considered a counterclockwise galaxy. If more than 30 peaks are detected in the radial intensity plot and the number of peaks shifting to the left is three times or more the number of peaks shifting to the right the galaxy is considered clockwise. If 30 peaks or more are detected and the number of peaks shifting to the right is three times or more the number of peaks shifting to the left the galaxy is considered clockwise. Otherwise, the handedness is undetermined. A detailed description of the Ganalyzer algorithm and its performance evaluation is provided in \citep{shamir2011ganalyzer,shamir2012handedness,dojcsak2014quantitative,hoehn2014characteristics}.

The 740,908 spiral galaxies were classified by their handedness as described above, providing a dataset with 82,242 clockwise galaxies and 80,272 counterclockwise galaxies. The remaining galaxies were not assigned with handedness and were therefore excluded from the analysis. The higher number of clockwise galaxies is aligned with previous experiments \citep{sha16,shamir2012handedness,hoehn2014characteristics}. Using cumulative binomial probability, the chance to have 82,242 or more successes in 162,514 trails is $P\simeq5\cdot10^{-7}$, when the probability of success is 0.5, showing that the  asymmetry in the number of clockwise and counterclockwise galaxies is statistically significant.

To test the consistency of the data, 400 galaxies classified by Ganalyzer as clockwise and 400 galaxies classified as counterclockwise were randomly separated from the dataset and inspected manually. Of the 800 galaxies 45 did not have a clear identifiable handedness, but none of the galaxies that were inspected was clearly assigned with the wrong handedness.

\begin{figure*}[ht]
\centering
\includegraphics[scale=0.5]{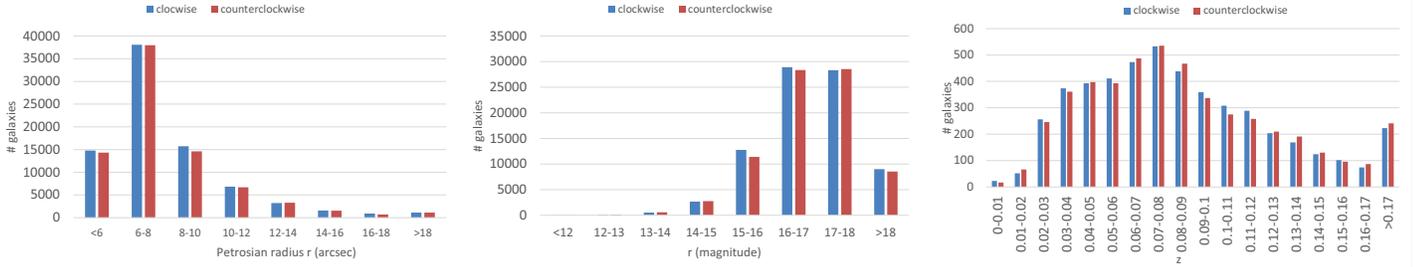}
\caption{Distribution of the r magnitude, Petrosian radius measured in the r band, and the distribution of redshift. The distribution of magnitude and radius was measured with the entire dataset of 162,514 galaxies. The distribution of the z is measured in a subset of 10,281 galaxies.}
\label{distribution}
\end{figure*}

Some of the photometric values in SDSS can be values such as -9999 or -1000. These values are flags and not actual photometric measurements, and were therefore removed from the analysis. The distribution of the r magnitude, Petrosian radius measured in the r band, and the redshift are displayed in Figure~\ref{distribution}. Most galaxies in the \citep{kum16} catalogue do not have spectra, and therefore just the subset of 10,281 galaxies with spectra were used to deduce the distribution of redshift.

\section{Results}
\label{results}

The mean De Vaucouleurs u-g, g-r, r-i, and i-z of clockwise and counterclockwise SDSS galaxies are shown in Table~\ref{dev_mag}. The table also shows the  two-tailed t-test statistical significance of the difference. Because four hypotheses are being tested, the statistical significance of each test needs to be corrected to avoid false positives. The Bonferroni correction \citep{goeman2014multiple} was applied to the t-test P values, so that the P value of each test is corrected for the total number of hypotheses analyzed in the experiment.

\begin{table*}[ht]
\caption{Mean, standard error of the mean, and two-tailed t-test P values of De Vaucouleurs u-g, g-r, r-i, and i-z of clockwise and counterclockwise galaxies.}
\label{dev_mag}
\begin{center}
{
 
\begin{tabular}{lcccc}

\hline
Colour       & Mean          & Mean                      & t-test    & Bonferroni-corrected \\
              &  clockwise    &  counterclockwise    &    P       & t-test P   \\   
\hline
u-g  & 1.1963$\pm$0.002 & 1.1948$\pm$0.002 & 0.61 & 1 \\
g-r  & 0.6173$\pm$0.0009 & 0.6176$\pm$0.001 & 0.8 & 1 \\
r-i  & 0.3299$\pm$0.0007 & 0.3266$\pm$0.0008 & 0.002 & 0.008 \\
i-z & 0.2136$\pm$0.001 & 0.2112$\pm$0.001 & 0.17 & 0.67 \\
\hline
\end{tabular}
}
\end{center}
\end{table*}

The table shows a statistically significant difference on the r-i colour, meaning that in galaxies with clockwise handedness the near infrared i band is more luminous than the r band compared to counterclockwise galaxies. The difference is also statistically significant after applying the Bonferroni correction to adjust the P values to the number of tests. When separating the galaxies randomly into two groups, the means r-i of the group is nearly identical ($\sim$0.3285) and the non-corrected P value is $\sim$0.8472.

In addition to random separation into two group, to ensure that no systematic error has affected the results, the experiment was repeated such that all galaxies were flipped horizontally using ImageMagick, and the experiment was repeated using the flipped galaxies. The results of that experiment are summarized in Table~\ref{flipped}.

\begin{table*}[ht]
\caption{Mean, standard error of the mean, and two-tailed t-test P values of De Vaucouleurs u-g, g-r, r-i, and i-z when all galaxies were mirrored.}
\label{flipped}
\begin{center}
{
\begin{tabular}{lcccc}
\hline
Colour       & Mean          & Mean                      & t-test    & Bonferroni-corrected \\
              &  clockwise    &  counterclockwise    &    P       & t-test P   \\   
\hline
u-g  & 1.1948$\pm$0.002 & 1.1963$\pm$0.002  & 0.61 & 1 \\
g-r  & 0.6176$\pm$0.001  & 0.6173$\pm$0.0009 & 0.8 & 1 \\
r-i  & 0.3266$\pm$0.0008 & 0.3299$\pm$0.0007 & 0.002 & 0.008 \\
i-z & 0.2112$\pm$0.001 & 0.2136$\pm$0.001 & 0.17 & 0.67 \\
\hline
\end{tabular}
}
\end{center}
\end{table*}

As the table shows, flipping the galaxies led to exactly the inverse results compared to the experiment when using the original galaxy images. The numbers of clockwise and counterclockwise galaxies are also inversed, and are 80,272 and 82,242, respectively. These identical results are expected since the Ganalyzer algorithm works by analyzing the radial intensity plot, and is therefore rotationally invariant.

Another common model to measure galaxy colours is the exponential magnitude model. Table~\ref{model_mag} shows the colours of the same galaxies measured using the exponential model magnitude.

\begin{table*}[ht]
\caption{Mean, standard error of the mean, and two-tailed t-test P values of exponential magnitude u-g, g-r, r-i, and i-z of clockwise and counterclockwise galaxies.}
\label{dev_mag}
\begin{center}
{
\begin{tabular}{lcccc}
\hline
Colour       & Mean              & Mean                            & t-test    & Bonferroni-corrected \\
              &   clockwise      &  counterclockwise           &    P       & t-test P   \\   
\hline
u-g  & 1.312$\pm$0.002 & 1.307$\pm$0.002 & 0.1 & 0.39 \\
g-r  & 0.615$\pm$0.0008 & 0.616$\pm$0.001 & 0.8 & 1 \\
r-i  & 0.331$\pm$0.0006 & 0.328$\pm$0.0007 & 0.003 & 0.013 \\
i-z & 0.178$\pm$0.001 & 0.176$\pm$0.001 & 0.21 & 0.86 \\
\hline
\end{tabular}
}
\end{center}
\end{table*}

Although the statistical significance is somewhat weaker, the exponential model magnitude also shows a statistically significant difference between clockwise and counterclockwise galaxies. Table~\ref{model_mag} shows the differences in the model magnitude, also showing statistically significant difference in the r-i colour of clockwise and counterclockwise galaxies. 

\begin{table*}[ht]
\caption{Mean, standard error of the mean, and two-tailed t-test P values of model magnitude u-g, g-r, r-i, and i-z of clockwise and counterclockwise galaxies.}
\label{model_mag}
\begin{center}
{
\begin{tabular}{lcccc}
\hline
Colour       & Mean              & Mean                            & t-test    & Bonferroni-corrected \\
              &   clockwise      &  counterclockwise           &    P       & t-test P   \\   
\hline
u-g  & 1.263$\pm$0.002 & 1.261$\pm$0.002 & 0.67 & 1 \\
g-r  & 0.648$\pm$0.0009 & 0.65$\pm$0.001 & 0.12 & 0.5 \\
r-i  & 0.355$\pm$0.0006 & 0.352$\pm$0.0007 & 0.006 & 0.03 \\
i-z & 0.227$\pm$0.001 & 0.226$\pm$0.001 & 0.36 & 1 \\
\hline
\end{tabular}
}
\end{center}
\end{table*}

\begin{figure}
\includegraphics[scale=0.95]{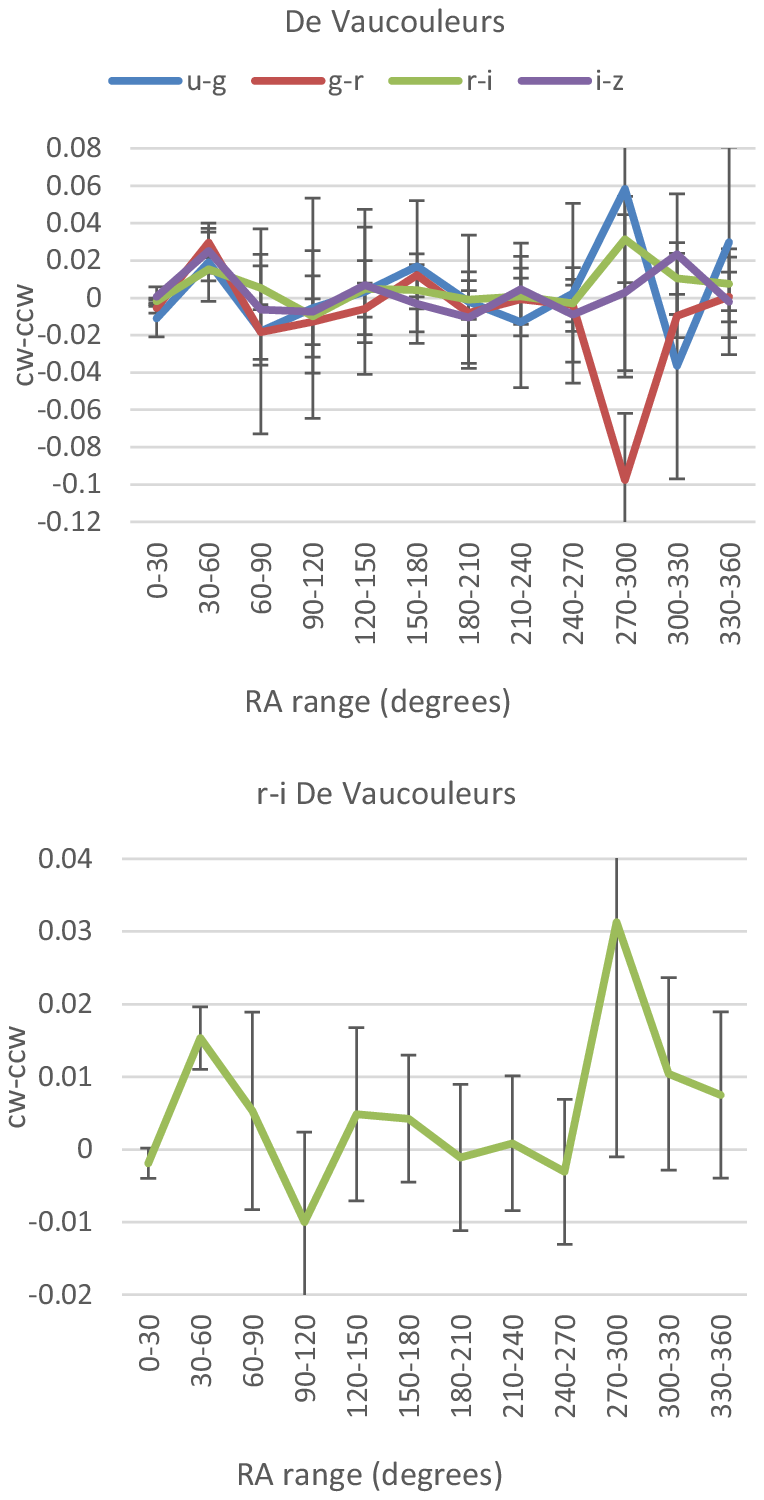}
\caption{Colour differences between u-g, g-r, r-i, and i-z of clockwise and counterclockwise galaxies in different RA ranges. The bottom panel shows the differences on the r-i colour. The error bars show the standard error of the mean.}
\label{dev_ra}
\end{figure}
   
Figure~\ref{dev_ra} shows the difference between the colours in different right ascension ranges. The figure shows different colour differences in different RA ranges. In the RA range of $(180^o,210^o)$ the colours of clockwise and counterclockwise galaxies are nearly the same, while the RA range of $(0^o,30^o)$ exhibits the strongest colour differences. The RA range $(270^o,300^o)$ shows stronger asymmetry, but the number of galaxies in that part of the sky is just 809, leading to high standard error. Table~\ref{dev_mag_30_60} shows the colour differences in that range. As can be learned from the table, the g-r, r-i, and i-z show very strong statistically significant difference between the colours in these bands.

\begin{table*}[ht]
\caption{Mean, standard error of the mean, and two-tailed t-test P values of De Vaucouleurs model magnitude u-g, g-r, r-i, and i-z of clockwise and counterclockwise galaxies in the RA range of 30$^o$-60$^o$.}
\label{dev_mag_30_60}
\begin{center}
{
\begin{tabular}{lcccc}
\hline
Colour       & Mean          & Mean                      & t-test    & Bonferroni-corrected \\
              &  clockwise     & counterclockwise    &    P       & t-test P   \\   
\hline
u-g  & 1.219$\pm$0.006 & 1.2$\pm$0.007 & 0.1 & 0.39 \\
g-r  & 0.659$\pm$0.003 & 0.629$\pm$0.003 & $<10^{-5}$ & $<10^{-5}$ \\
r-i  & 0.351$\pm$0.002 & 0.335$\pm$0.002 & $<10^{-5}$ & $<10^{-5}$ \\
i-z & 0.24$\pm$0.003 & 0.214$\pm$0.004 & $<10^{-5}$ & $<10^{-5}$ \\
\hline
\end{tabular}
}
\end{center}
\end{table*}

Another statistically significant difference is the g-r colour, measured in the RA range $(150^o-180^o)$. The mean g-r of counterclockwise galaxies in that RA range is $\sim$0.5789, and for clockwise galaxies the mean g-r is $\sim$0.591. The two-tailed t-test probability that these two means are not different is P$<0.0004$.

\begin{figure*}
\centering
\includegraphics[scale=1.0]{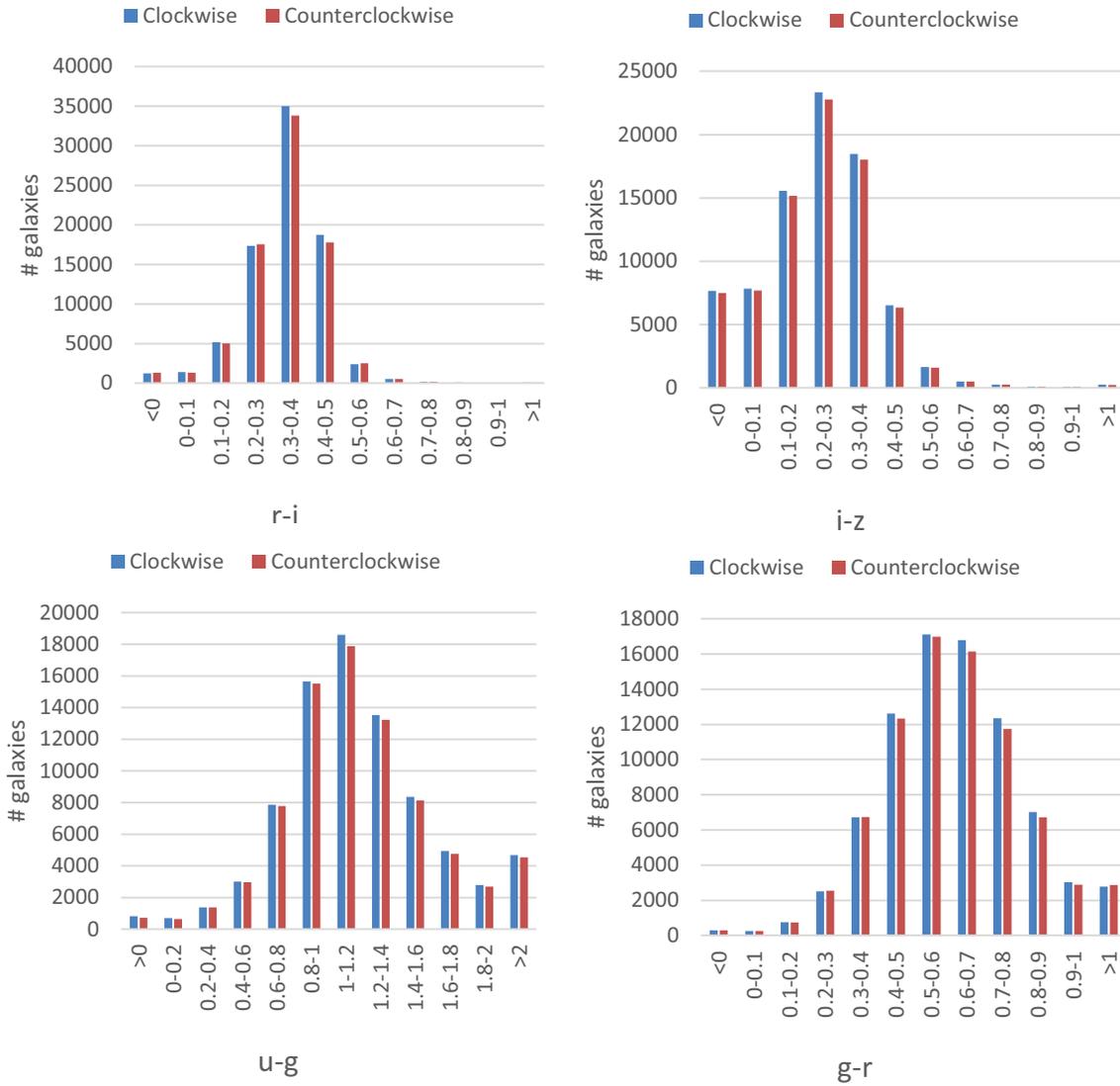}
\caption{Histogram of the distribution of the different colours}
\label{histograms}
\end{figure*}

Figure~\ref{histograms} shows the histogram of the distribution of the different colours among clockwise and counterclockwise galaxies. The histograms show that the colour differences are not distributed uniformly. For instance, in the r-i colour, the range 0.3 to 0.5 has more clockwise galaxies, while the number of counterclockwise galaxies is higher when the r-i is between 0.2 and 0.3.

\section{Discussion}
\label{discussion}

While spiral galaxies have been attracting substantial research attention since their discovery in 1773, the physics of spiral arms is still an open fundamental question \citep{dobbs2014dawes,griv2014density}. Spiral galaxies can be separated into several different types \citep{elmegreen1982flocculent,elmegreen1987arm}, and the formation of spiral arms has been linked to quasi-stationary density waves \citep{lin1964spiral,bertin1993role,elmegreen1993grand}, dark matter halos \citep{khoperskov2013interaction,tutukov2006role,chang2011dark,del2016dark}, stochastic star formation \citep{mueller1976propagating,gerola1978stochastic}, spiral instabilities \citep{sellwood1984spiral,sellwood2000spiral,baba2013dynamics}, radial transport of dust \citep{vorobyov2006radial}, and tidal interactions \citep{toomre1972galactic,meidt2013gas,mcconnachie2009remnants}.

The colour of a galaxy clearly provides important information about its physical characteristics, composition, and stellar profile. Previous studies using populations of galaxies have shown dependence between colour and morphology \citep{de1961integrated}, intrinsic luminosities \citep{fioc1999statistical}, star formation \citep{ferreras1999segregated,tojeiro2013different}, clustering \citep{brown2000clustering,brown2003red,coil2004evolution}, dust grain population \citep{shalima2015dust}, and the radial profile \citep{strateva2001color}.

The results of this study show that in SDSS there is a statistically significant link between the colour of a spiral galaxy and its handedness, with the strongest difference observed in the r-i colour. The results also show that the colours are different in different right ascension ranges, with the most substantial difference in the RA range $(30^o,60^o)$. However, the differences in the r-i colour are observed in a population of galaxies that covers a large part of the sky, much larger than any known superstructure of gravitationally interacting galaxies.


The analysis applied in this study compares the colour of SDSS clockwise and anti-clockwise galaxies in the same sky regions. Although the asymmetry can be the results of a measurement error, it is difficult to identify reasons for such differences due to errors in SDSS pipeline, as both clockwise and anti-clockwise galaxies were measured in the same parts of the sky, and were separated into two classes only after the images and photometric measurements were acquired. The differences observed in this study can also be explained by a violation of the cosmological principle \citep{longo2011detection,shamir2012handedness,horsch2013,gullu2014spin,tasseten2016,chechin2016rotation}, as the differences in asymmetry in different directions of observation might indicate that the local universe is not isotropic. While several observations have shown structures that violate the homogeneity assumption of the cosmological principle \citep{balazs2015giant}, the observation reported here provides preliminary evidence for the violation also of the isotropy assumption.

\section*{Acknowledgments}

The research was supported in part by National Science Foundation grant IIS-1546079. I would like to thank the reviewer, Anze Slosar, for the insightful comments and ideas that helped to improve the manuscript.

\bibliographystyle{spr-mp-nameyear-cnd}

\bibliography{color_assym}

\end{document}